\documentclass[Physsubmission, Phys]{SciPost}

\hypersetup{
    colorlinks,
    linkcolor={red!50!black},
    citecolor={blue!50!black},
    urlcolor={blue!80!black}
}

\usepackage[bitstream-charter]{mathdesign}
\DeclareSymbolFont{usualmathcal}{OMS}{cmsy}{m}{n}
\DeclareSymbolFontAlphabet{\mathcal}{usualmathcal}

\usepackage[printonlyused]{acronym}
\usepackage{mathtools}
\usepackage{slashed}
\usepackage{nicefrac}
\usepackage[capitalize]{cleveref}
\usepackage[position=b, labelformat=simple]{subcaption}

\newcommand{\citere}[1]{Ref.~\cite{#1}}
\newcommand{\citeres}[1]{Refs.~\cite{#1}}

\newcommand{\dd}{\mathrm{d}}
\newcommand{\ii}{\mathrm{i}}
\newcommand{\EulerGamma}{\gamma_\mathrm{E}}
\newcommand{\order}[1]{\mathcal{O}(#1)}
\newcommand{\deriv}[3]{\frac{\partial\ifthenelse{\equal{#1}{}}{}{^{#1}} #2}{\partial #3\ifthenelse{\equal{#1}{}}{}{^{#1}}}}
\newcommand{\dderiv}[3]{\frac{\dd\ifthenelse{\equal{#1}{}}{}{^{#1}} #2}{\dd #3\ifthenelse{\equal{#1}{}}{}{^{#1}}}}
\newcommand{\calo}{\mathcal{O}}
\newcommand{\tcalo}{\tilde{\calo}}
\newcommand{\bcalo}{\bar{\calo}}
\newcommand{\alphas}{\alpha_\mathrm{s}}
\newcommand{\msbar}{\ensuremath{\overline{\mathrm{MS}}}}
\newcommand{\tr}{T_\mathrm{R}}
\newcommand{\nc}{N_\mathrm{C}}
\newcommand{\na}{N_\mathrm{A}}
\newcommand{\cf}{C_\mathrm{F}}
\newcommand{\ca}{C_\mathrm{A}}
\newcommand{\nf}{n_\mathrm{f}}
\newcommand{\nh}{n_\mathrm{h}}

\newcommand{\tf}{\tr\nf}

\newcommand{\ren}{\mathrm{R}}
\newcommand{\bare}{\mathrm{B}}

\newcommand{\betagf}[1]{\hat{\beta}_{\rho\ifthenelse{\equal{#1}{}}{}{, }#1}}

\newcounter{notecount}

\makeatletter
\define@key{Gin}{trimplots}[]{%
  \begingroup\edef\x{%
    \endgroup\noexpand\setkeys{Gin}{trim = 3 12 9 6, clip}%
  }\x
}

\acrodef{emt}[EMT]{\emph{energy-momentum tensor}}
\acrodef{lo}[LO]{\emph{leading order}}
\acrodef{nlo}[NLO]{\emph{next-to-leading order}}
\acrodef{nnlo}[NNLO]{\emph{next-to-next-to-leading order}}
\acrodef{ope}[OPE]{\emph{operator product expansion}}
\acrodef{qcd}[QCD]{\emph{Quantum Chromodynamics}}
\acrodef{rg}[RG]{\emph{renormalization group}}
\acrodef{uv}[UV]{\emph{ultra-violet}}
\acrodef{vev}[VEV]{\emph{vacuum expectation value}}
\acrodef{vpf}[VPF]{\emph{vacuum polarization function}}

\begin{document}

\begin{center}{\Large \textbf{
Applications of the Perturbative Gradient Flow\\
}}\end{center}

\begin{center}
Fabian Lange\textsuperscript{1,2$\star$}
\end{center}

\begin{center}
{\bf 1} Institut f\"ur Theoretische Teilchenphysik, Karlsruhe Institute of Technology (KIT), Wolfgang-Gaede Straße 1, 76128 Karlsruhe, Germany
\\
{\bf 2} Institut f\"ur Astroteilchenphysik, Karlsruhe Institute of Technology (KIT), Hermann-von-Helmholtz-Platz 1, 76344 Eggenstein-Leopoldshafen, Germany
\\
* fabian.lange@kit.edu
\end{center}

\begin{center}
\today
\end{center}


\definecolor{palegray}{gray}{0.95}
\begin{center}
\colorbox{palegray}{
  \begin{tabular}{rr}
  \begin{minipage}{0.1\textwidth}
    \includegraphics[width=35mm]{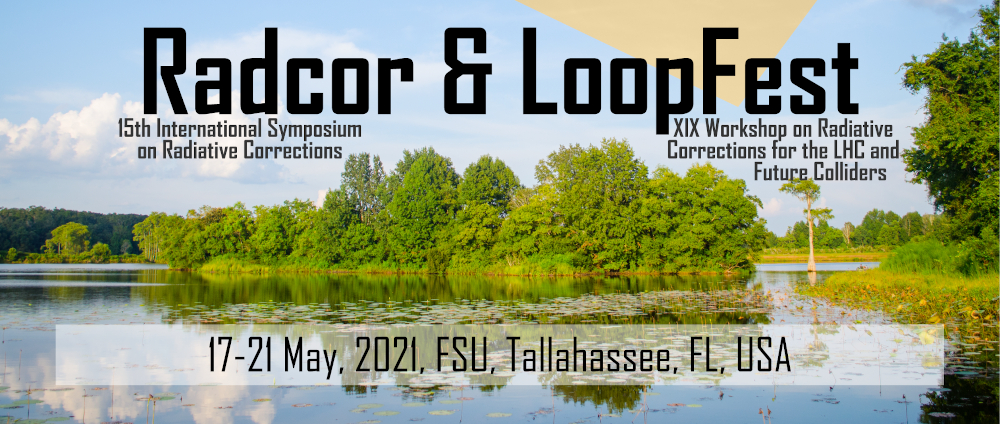}
  \end{minipage}
  &
  \begin{minipage}{0.85\textwidth}
    \begin{center}
    {\it 15th International Symposium on Radiative Corrections: \\Applications of Quantum Field Theory to Phenomenology,}\\
    {\it FSU, Tallahasse, FL, USA, 17-21 May 2021} \\
    \doi{10.21468/SciPostPhysProc.?}\\
    \end{center}
  \end{minipage}
\end{tabular}
}
\end{center}

\section*{Abstract}
{\bf
Over the last decade the gradient flow formalism became an important tool for lattice simulations of Quantum Chromodynamics.
It offers remarkable renormalization properties which pave the way for cross-fertilization between perturbative and lattice calculations.
In this contribution we discuss the perturbative approach.
As first application we compute vacuum expectation values of flowed operators which could help to extract parameters like the strong coupling constant from lattice simulations.
Afterwards, we apply the flowed operator product expansion to the time-ordered product of two currents which could be employed for an alternative first-principle evaluation of vacuum polarization functions on the lattice.
}

\vspace{10pt}
\noindent\rule{\textwidth}{1pt}
\tableofcontents\thispagestyle{fancy}
\noindent\rule{\textwidth}{1pt}
\vspace{10pt}

\section{Introduction}
\label{sec:intro}

The gradient flow formalism introduced in \citeres{Narayanan:2006rf,Luscher:2009eq,Luscher:2010iy} became an important tool for simulations of \ac{qcd} on the lattice over the last decade.
Most prominently, it led to new strategies to set the scale of the lattice, see e.g.\ \citeres{Luscher:2010iy,Borsanyi:2012zs,Sommer:2014mea}.
Moreover, it provides strong renormalization properties.
Especially, \emph{flowed}\footnote{
  We use the terms \emph{flowed} and \emph{regular} to distinguish quantities defined at flow time $t>0$ from those defined at $t=0$.
}
composite operators constructed from \emph{flowed} fields do not require renormalization~\cite{Luscher:2011bx}.
Therefore, they do not mix under \ac{rg} running which allows one to match results of lattice and perturbative calculations without scheme transformation.
One prominent application is the extraction of the strong coupling $\alphas$ from lattice simulations which, however, did not yield competitive results yet, see \citere{DallaBrida:2020pag} for a recent review.

A powerful tool is the so-called \emph{small-flow-time expansion} which leads to a relation between flowed and regular operators related by a flow-time dependent mixing matrix~\cite{Luscher:2011bx}.
By inverting the mixing matrix, one obtains a flowed \ac{ope} which expresses the regular operators through the corresponding flowed operators\cite{Suzuki:2013gza,Makino:2014taa,Monahan:2015lha}.
This was first utilized to construct a regularization independent formula for the \ac{emt} of \ac{qcd}~\cite{Suzuki:2013gza,Makino:2014taa}.

In this contribution we briefly introduce the perturbative treatment of the gradient flow at infinite volume\footnote{
  At finite volume different techniques are required, see e.g.\ \citere{DallaBrida:2017tru}.
}
in \cref{sec:basics} which allows us to compute \acp{vev} of flowed operators through \ac{nnlo} in \cref{sec:VEVs}.
In \cref{sec:flope}, we discuss the flowed \ac{ope} and apply it to \acp{vpf} which might pave the way for an alternative determination of the hadronic corrections to the anomalous magnetic moment and other observables.

\section{Perturbative Gradient Flow}
\label{sec:basics}

The gradient flow formalism continues the gluon and quark fields $A^a_\mu(x)$ and $\psi(x)$ of regular \ac{qcd} from $D=4-2\epsilon$ Euclidean dimensions to the fields $B^a_\mu(t,x)$ and $\chi(t,x$) additionally depending on the \emph{flow time} $t > 0$ through the boundary conditions
\begin{equation}
  \begin{split}
    B_\mu^a (t=0,x) = A_\mu^a (x)\,,\qquad \chi (t=0,x) = \psi(x)
    \label{eq:boundary-conditions}
  \end{split}
\end{equation}
and the flow equations~\cite{Luscher:2010iy,Luscher:2013cpa}
\begin{equation}
  \partial_t B_\mu^a = \mathcal{D}^{ab}_\nu G_{\nu\mu}^b + \kappa \mathcal{D}^{ab}_\mu \partial_\nu B_\nu^b ,\qquad
  \partial_t \chi = \Delta \chi - \kappa \partial_\mu B_\mu^a T^a \chi ,\qquad
  \partial_t \bar \chi = \bar \chi \overleftarrow \Delta + \kappa \bar \chi \partial_\mu B_\mu^a T^a ,
  \label{eq:flow-equations}
\end{equation}
where
\begin{equation}
  \begin{gathered}
    G_{\mu\nu}^a = \partial_\mu B_\nu^a - \partial_\nu B_\mu^a + f^{abc} B_\mu^b B_\nu^c , \qquad \mathcal{D}_\mu^{ab} = \delta^{ab} \partial_\mu - f^{abc} B_\mu^c , \\
    \Delta = \mathcal{D}^\mathrm{F}_\mu \mathcal{D}^\mathrm{F}_\mu ,\qquad \overleftarrow{\Delta} = \overleftarrow{\mathcal{D}}^\mathrm{F}_\mu \overleftarrow{\mathcal{D}}^\mathrm{F}_\mu , \qquad \mathcal{D}^\mathrm{F}_\mu = \partial_\mu + B^a_\mu T^a ,\qquad \overleftarrow{\mathcal{D}}^\mathrm{F}_\mu = \overleftarrow{\partial}_\mu - B^a_\mu T^a .
  \end{gathered}
\end{equation}
The flow time $t$ is a parameter of mass dimension minus two and we use the short-hand notation $\partial_t \equiv \frac{\partial}{\partial t}$.
The symmetry generators $T^a$ in the fundamental representation and the structure constants $f^{abc}$ are defined through
\begin{equation}
  \label{eq:generator-commutator}
  [T^a,T^b] = f^{abc}T^c , \qquad \mathrm{Tr}(T^a T^b) = - \tr \delta^{ab} .
\end{equation}
$\kappa$ is an additional gauge parameter and all physical observables should be independent of it~\cite{Luscher:2010iy}.
In perturbative calculations it is usually most convenient to set $\kappa = 1$.

The flow equations (\ref{eq:flow-equations}) can be incorporated into a Lagrangian formalism by defining
\begin{equation}
  \mathcal{L} = \mathcal{L}_\mathrm{QCD} + \mathcal{L}_\textrm{gauge-fixing} + \mathcal{L}_\mathrm{ghost} + \mathcal{L}_{B} + \mathcal{L}_{\chi} .
  \label{eq:Lagrangian_complete}
\end{equation}
The first three terms constituting the regular Yang-Mills Lagrangian are given by
\begin{equation}
  \begin{gathered}
    \mathcal{L}_\mathrm{QCD} = \frac{1}{4g_\bare^2} F_{\mu \nu}^a F_{\mu\nu}^a + \sum_{f=1}^{\nf}\bar{\psi}_f ( \slashed{D}^\mathrm{F}+m_{f,\bare}) \psi_f ,\\
    \mathcal{L}_\textrm{gauge-fixing} = \frac{1}{2 g_\bare^2 \xi}(\partial_\mu A_\mu^a)^2 ,\qquad
    \mathcal{L}_\mathrm{ghost} = \frac{1}{g_\bare^2}\partial_\mu \bar{c}^a D_\mu^{ab} c^b ,
    \label{eq:Lagrangian-parts}
  \end{gathered}
\end{equation}
where
\begin{equation}
  F_{\mu\nu}^a = \partial_\mu A^a_\nu - \partial_\nu A^a_\mu + f^{abc}A_\mu^bA_\nu^c , \qquad D_\mu^\mathrm{F} = \partial_\mu + A_\mu^a T^a ,\qquad D_\mu^{ab} = \delta^{ab} \partial_\mu - f^{abc} A_\mu^c .
\end{equation}
The flow equations are incorporated through
\begin{align}
  \label{eq:Lagrangian_flow}
  \mathcal{L}_{B} &= -2 \int_0^\infty \mathrm{d}t \, \textrm{Tr} \left[ L_\mu^a T^a \left(\partial_t B_\mu^b T^b -\mathcal{D}_\nu^{bc} G_{\nu \mu}^c T^b - \kappa \mathcal{D}_\mu^{bc} \partial_\nu B_\nu^c T^b \right) \right] ,\\
  \mathcal{L}_{\chi} &= \sum_{f=1}^{\nf}\int_0^\infty \mathrm{d}t \Big[ \bar{\lambda}_f \left(\partial_t - \Delta + \kappa \left(\partial_\mu B_\mu^a\right) T^a \right) \chi_f + \bar{\chi}_f \left(\overleftarrow{\partial_t} - \overleftarrow{\Delta} - \kappa \left(\partial_\mu B_\mu^a\right) T^a\right) \lambda_f \Big] , \nonumber
\end{align}
where $L_\mu^a(t,x)$, $\lambda_f(t,x)$, and $\bar\lambda_f(t,x)$ are Lagrange multiplier fields~\cite{Luscher:2011bx,Luscher:2013cpa}.
Their Euler-Lagrange equations lead to \cref{eq:flow-equations}.

The Feynman rules for perturbative calculations can be derived from the Lagrangian employing standard techniques~\cite{Luscher:2011bx,Luscher:2013cpa} and the complete list can be found in \citere{Artz:2019bpr}.

\section{Vacuum Expectation Values}
\label{sec:VEVs}

\acp{vev} of gauge-invariant operators at finite flow time are among the simplest quantities one can consider within the gradient flow formalism.
As mentioned before, these operators do not require any \ac{uv} renormalization beyond that of regular \ac{qcd} and that of the involved flowed fields~\cite{Luscher:2011bx}.
This means that the operators do not mix under \ac{rg} running, which makes it particularly simple to combine results from different regularization schemes.

The renormalization of the coupling and the masses follows the usual prescription with the known \ac{qcd} renormalization constants.
Throughout this contribution we employ the \msbar{} scheme and refer to \citeres{Artz:2019bpr,Harlander:2020duo} for details.

The flowed gauge field $B^a_\mu(t,x)$ does not require renormalization so that matrix elements of the gluon action density
\begin{align}
  E (t,x) \equiv \tfrac{1}{4} G_{\mu \nu}^a(t,x) G_{\mu \nu}^a(t,x)
  \label{eq:def_E}
\end{align}
are finite after just the renormalization of $g$ and $m_f$~\cite{Luscher:2010iy,Luscher:2011bx}.
Hence, a direct comparison of results obtained in different regularization schemes is possible.

In contrast, flowed quark fields require a renormalization factor $Z_\chi^{\nicefrac{1}{2}}(\alphas)$ in order to render Green's functions finite.
In the \msbar{} scheme it reads
\begin{equation}
  Z^{-1}_\chi(\alphas) = 1 -\frac{\alphas}{4\pi}\frac{\gamma_{\chi,0}}{\epsilon} + \left(\frac{\alphas}{4\pi}\right)^2\left[\frac{\gamma_{\chi,0}}{2\epsilon^2}\left(\gamma_{\chi,0} + \beta_0\right) - \frac{\gamma_{\chi,1}}{2\epsilon}\right] + \order{\alphas^3} ,
  \label{eq:Zchi}
\end{equation}
with
\begin{equation}
  \gamma_{\chi,0} = 3\,\cf , \qquad \gamma_{\chi, 1} = \left( \tfrac{223}{6} - 8 \ln 2 \right) \ca \cf - \left( \tfrac{3}{2} + 8 \ln 2 \right) \cf^2 - \tfrac{22}{3} \cf \tf .
\end{equation}
$\gamma_{\chi,0}$ has been computed in \citere{Luscher:2013cpa}, whereas $\gamma_{\chi, 1}$ has been obtained by requiring that the \ac{nnlo} calculations in \citeres{Harlander:2018zpi,Artz:2019bpr} become finite.

The scalar quark density
\begin{align}
  S (t,x) \equiv Z_\chi\,\sum_{f=1}^{\nf}\bar{\chi}_f(t,x) \chi_f(t,x)
  \label{eq:def_S}
\end{align}
thus acquires an anomalous dimension, which prevents a direct comparison of results from different regularization schemes.
This can be cured by working with \emph{ringed quark fields}~\cite{Makino:2014taa}, which amounts to renormalizing the flowed quark fields with
\begin{equation}
  \mathring{Z}_\chi(t,\mu) = -\frac{2\nc \nf}{(4\pi t)^2} \cdot \frac{1}{\left.\langle R(t)\rangle\right|_{m=0}}
  \label{eq:Zchiring}
\end{equation}
instead of $Z_\chi$, where
\begin{equation}
  R(t,x) = \sum_{f=1}^{\nf}\bar{\chi}_f(t,x) \overleftrightarrow{\slashed{\mathcal{D}}}^\mathrm{F} \chi_f(t,x) \qquad \text{with} \qquad \overleftrightarrow{\mathcal{D}}^\mathrm{F}_\mu = \mathcal{D}^\mathrm{F}_\mu - \overleftarrow{\mathcal{D}}^\mathrm{F}_\mu
\end{equation}
is the quark kinetic operator.
This corresponds to a ``physical'' renormalization scheme, which means that the anomalous dimension of the operator
\begin{equation}
  \mathring{S}(t,x) = \zeta_\chi(t,\mu)\,S(t,x) \qquad\mbox{with}\qquad \zeta_\chi(t,\mu) \equiv Z_\chi^{-1}\mathring{Z}_\chi(t,\mu)
  \label{eq:zetachi}
\end{equation}
vanishes.

\begin{figure}
  \centering
  \begin{subfigure}{0.3\textwidth}
    \centering
    \includegraphics[scale=0.9]{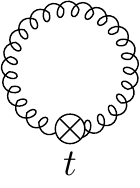}
    \caption{}
  \end{subfigure}
  \hfil
  \begin{subfigure}{0.3\textwidth}
    \centering
    \includegraphics[scale=0.9]{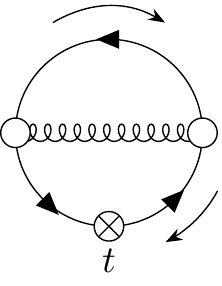}
    \caption{}
  \end{subfigure}
  \hfil
  \begin{subfigure}{0.3\textwidth}
    \centering
    \includegraphics[scale=0.9]{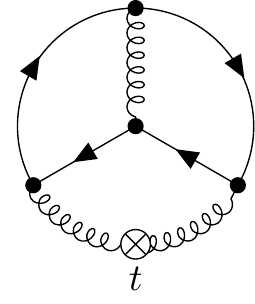}
    \caption{}
  \end{subfigure}
  \hfil
  \begin{subfigure}{0.3\textwidth}
    \centering
    \includegraphics[scale=0.9]{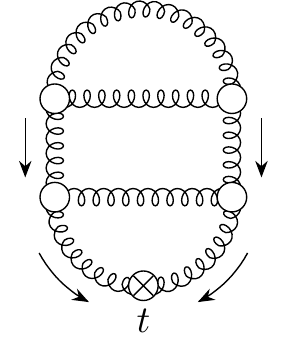}
    \caption{}
  \end{subfigure}
  \hfil
  \begin{subfigure}{0.3\textwidth}
    \centering
    \includegraphics[scale=0.9]{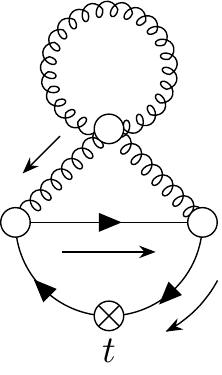}
    \caption{}
  \end{subfigure}
  \hfil
  \begin{subfigure}{0.3\textwidth}
    \centering
    \includegraphics[scale=0.9]{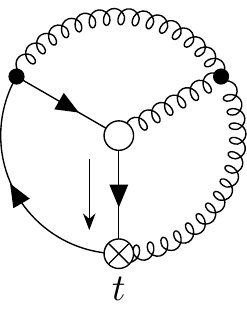}
    \caption{}
  \end{subfigure}
	\caption{Sample diagrams for the \acp{vev} through \ac{nnlo}. Produced with Ti\textit{k}Z-Feynman~\cite{Ellis:2016jkw}.}
	\label{fig:vevs-example_diagrams}
\end{figure}

The Feynman rules for the operators $E(t,x)$, $S(t,x)$, and $R(t,x)$ can again be derived by standard techniques and are listed in \citere{Artz:2019bpr}.
They result in Feynman diagrams like the samples shown in \cref{fig:vevs-example_diagrams}.
In \citere{Artz:2019bpr} we set up a program chain to automatically generate~\cite{Nogueira:1991ex,Nogueira:2006pq} and process~\cite{Harlander:1997zb,Seidensticker:1999bb,Vermaseren:2000nd,Kuipers:2012rf,vanRitbergen:1998pn} these diagrams as well as to perform a reduction to master integrals~\cite{Tkachov:1981wb,Chetyrkin:1981qh,Laporta:2001dd,Maierhoefer:2017hyi,Klappert:2020nbg,Klappert:2019emp,Klappert:2020aqs} and to solve those subsequently~\cite{Harlander:2016vzb,Binoth:2000ps,Binoth:2003ak,Smirnov:2008py,Smirnov:2009pb,Smirnov:2013eza,Genz:1983,Mathematica,Huber:2005yg,Huber:2007dx,Panzer:2014caa}.

For the gluon action density we then find
\begin{equation}
  \begin{gathered}
    \left. \langle E(t) \rangle \right|_{m=0} = \frac{3\alphas}{4\pi t^2} \frac{\na}{8} \left[1 + \frac{\alphas}{4\pi} e_1(\mu^2t) + \left(\frac{\alphas}{4\pi}\right)^2 e_2(\mu^2t) + \mathcal{O}(\alphas^3)\right] , \\
    e_1(z) = e_{1,0} + \beta_0 \, L(z) , \qquad e_2(z) = e_{2,0} + (2\beta_0\,e_{1,0}+\beta_1) \, L(z) + \beta_0^2 \, L^2(z) ,
  \end{gathered}
  \label{eq:e_nnlo}
\end{equation}
where $\alphas=\alphas(\mu)$, with $\mu$ the renormalization scale,
\begin{equation}
  L(z) \equiv \ln(2z) + \EulerGamma ,
  \label{eq:Lmut}
\end{equation}
and with the \msbar{} coefficients $\beta_0$, $\beta_1$.
For the non-logarithmic coefficients $e_{i,j}$ we find
\begin{align}
  \label{eq:econst}
  e_{0,0} &= 1 ,\qquad e_{1,0} = \big(\tfrac{52}{9} + \tfrac{22}{3} \ln 2 - 3 \ln 3\big) \ca - \tfrac{8}{9} \tf \\
  e_{2,0} &= 27.9784\, \ca^2  - (31.5652\ldots) \, \ca\tf + \big(16 \zeta(3) - \tfrac{43}{3}\big) \cf\tf + \big(\tfrac{8\pi^2}{27} - \tfrac{80}{81}\big) \tr^2\nf^2 , \nonumber
\end{align}
where $\zeta(3) = 1.20206\ldots$.
The \ac{nlo} coefficient $e_1$ was first evaluated in \citere{Luscher:2010iy} and the \ac{nnlo} coefficient in \citere{Harlander:2016vzb}.
The three dots in the coefficient of $\ca\tf$ indicate that we were able to obtain the expression in analytical form in \citere{Artz:2019bpr}.
Our estimate of the numerical accuracy for the $\ca^2$ coefficient is at least six digits beyond the four decimal places shown here.

\begin{figure}[ht]
  \centering
  \hfil
  \begin{subfigure}{0.49\textwidth}
    \centering
    \includegraphics[width=\textwidth, trimplots]{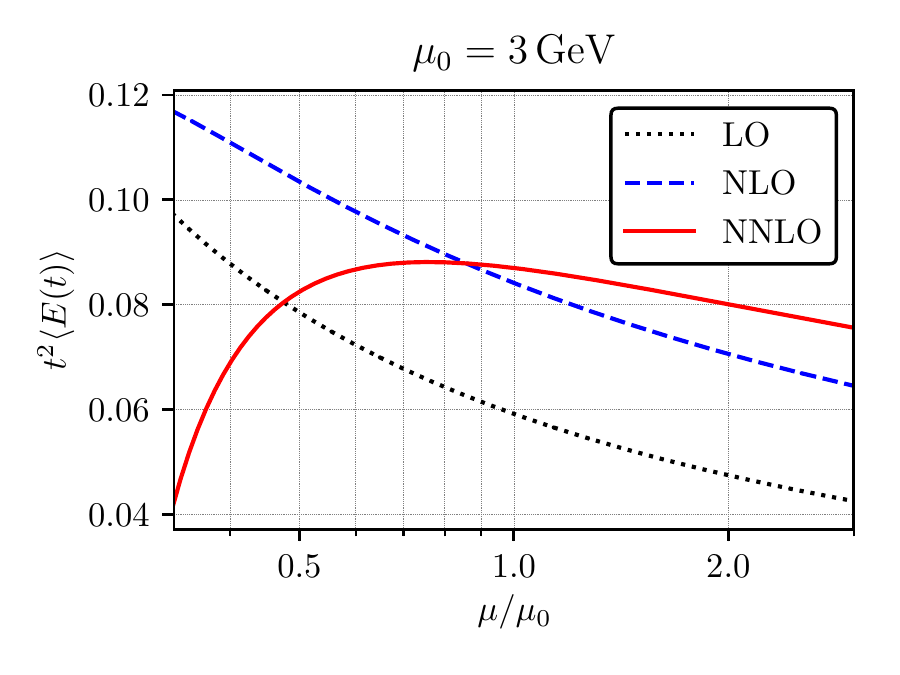}
    \caption{}
  \end{subfigure}
  \hfil
  \begin{subfigure}{0.49\textwidth}
    \centering
    \includegraphics[width=\textwidth, trimplots]{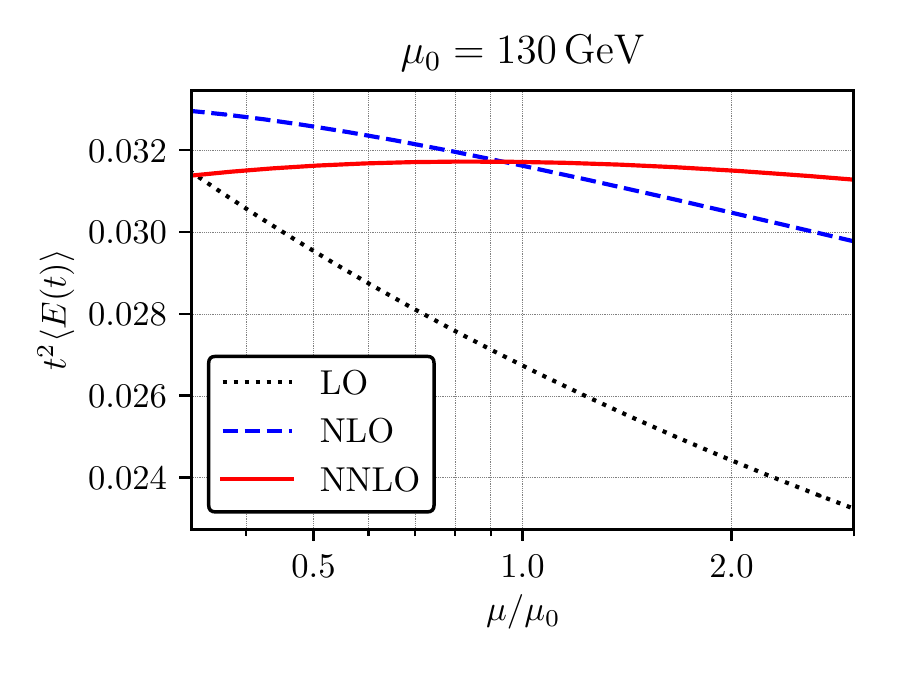}
    \caption{}
  \end{subfigure}
  \hfil
  \caption{Renormalization scale dependence of $t^2 \langle E(t) \rangle$ in \ac{qcd} for two different central scales $\mu_0=e^{-\EulerGamma/2}/\sqrt{2t}$.
  See \citere{Artz:2019bpr} for details.}
  \label{fig:scale-dependence}
\end{figure}

Since these \acp{vev} are formally independent of the renormalization scale $\mu$, the residual scale dependence can be used to study the behavior of the perturbative expansion.
As shown in \cref{fig:scale-dependence} for $\langle E(t) \rangle$, it is well behaved at high energies and still decent around a central scale of $3$\,GeV.
For a detailed discussion and results for $\langle\mathring{S}(t)\rangle$ and $\zeta_\chi(t,\mu)$ we refer to \citere{Artz:2019bpr}.

The proportionality of $\langle E(t) \rangle$ to $\alphas$ suggests to define a \emph{gradient flow coupling}
\begin{equation}
  \alpha_\text{GF} (t) \equiv \frac{1}{\mathcal{N}} \frac{32\pi}{3\na} t^2 \langle E(t)\rangle_\text{lattice} ,
\end{equation}
based on the determination of $\langle E(t)\rangle_\text{lattice}$ through lattice simulations, where $\mathcal{N}$ accounts for boundary conditions~\cite{Luscher:2014kea,Fodor:2017die,Hasenfratz:2019hpg,Fodor:2019ypi}.
Our perturbative result in \cref{eq:e_nnlo} then might help in the extrapolation to infinite volume.

\section{Flowed Operator Product Expansion}
\label{sec:flope}

A powerful concept in the gradient flow formalism is the flowed \ac{ope}~\cite{Suzuki:2013gza,Makino:2014taa,Monahan:2015lha}.
Consider a set of operators $\calo_{i}(x)$ and a corresponding set of flowed operators $\tcalo_{i}(t,x)$ which are constructed from flowed fields.
They are related by the \emph{small-flow-time expansion}
\begin{equation}
  \tcalo_{i}(t,x) = \sum_j \zeta_{ij}(t)\calo_{j}(x) + O(t)
  \label{eq:small-flow-time-expansion}
\end{equation}
with the flow-time dependent mixing matrix $\zeta_{ij}(t)$~\cite{Luscher:2011bx}.
By inverting \cref{eq:small-flow-time-expansion} one can then express any linear combination of the $\calo_{i}(x)$ through their flowed counterparts:
\begin{align}
  T = \sum_i C_i \calo_i = \sum_{i,j} C_i \zeta^{-1}_{ij} \tcalo_j + O(t) \equiv \sum_j \tilde C_j \tcalo_j + O(t),
\end{align}
where $C_i$ are the Wilson coefficients for the object $T$.
Since the flowed operators do not require renormalization beyond field and coupling renormalization~\cite{Luscher:2011bx}, the r.h.s.\ is scheme independent.
Thus, one can directly relate $T$ in different schemes, for example lattice and perturbative schemes, by employing the flowed \ac{ope}.
First, it was applied to the \ac{emt}~\cite{Suzuki:2013gza,Makino:2014taa,Harlander:2018zpi} which led to promising thermodynamical results~\cite{Asakawa:2013laa,Taniguchi:2016ofw,Kitazawa:2016dsl,Kitazawa:2017qab,Yanagihara:2018qqg,Iritani:2018idk,Kitazawa:2019otp,Taniguchi:2020mgg,Yanagihara:2020tvs}.
Other applications include charge conjugation parity violating operators for the nucleon electric dipole moment~\cite{Rizik:2020naq} or the electroweak Hamiltonian~\cite{Suzuki:2020zue}.
We now apply it to \acfp{vpf}.

\section{Hadronic Vacuum Polarization using Gradient Flow}
\label{sec:vacuum-polarization}

\acp{vpf} for (axial-)vector and (pseudo-)scalar particles are important objects in \ac{qcd}.
Through the optical theorem, their imaginary part is directly related to physical observables such as the decay rates of the $Z$- or the Higgs boson, or the hadronic R-ratio.
Moreover, \acp{vpf} also contribute indirectly to physical observables such as anomalous magnetic moments~\cite{Jegerlehner:2017gek,Aoyama:2020ynm}, the definition of short-distance quark masses~\cite{Chetyrkin:2010ic}, or hadronic contributions to the coupling of Quantum Electrodynamics~\cite{Crivellin:2020zul,Keshavarzi:2020bfy}.
However, the latter applications involve an integration of the \acp{vpf} over the non-perturbative regime.
They are typically computed from experimental data with the help of dispersion relations.
First-principle lattice calculations have started to become competitive with these dispersive approaches only very recently.
However, for the prominent topic of the hadronic vacuum polarization contribution to the muon's anomalous magnetic moment, the two approaches lead to incompatible results~\cite{Borsanyi:2020mff}.

The perturbative and non-perturbative regimes of \acp{vpf} can explicitly be demonstrated through the \ac{ope} (see, e.g., \citere{Dominguez:2014vca}):
\begin{equation}
  \begin{split}
    T(Q) \equiv \int\dd^4 x\, e^{\ii Qx} \langle Tj(x)j(0)\rangle \stackrel{Q^2\to \infty}{\sim} \sum_{k,n} C^{(k),\bare}_{n}(Q)\langle\calo^{(k)}_{n}(x=0)\rangle ,
  \end{split}
  \label{eq:ccope}
\end{equation}
where $j(x)$ generically stands for a scalar, pseudo-scalar, vector, axial-vector, or tensor current, and $k$ labels the mass dimension.
In principle, the coefficients $C_{n}^{(k),\bare}$ on the r.h.s.\ of \cref{eq:ccope} depend on the quantum numbers of the currents, but we suppress such indices in the following.
We furthermore assume that possible global divergences are subtracted off of $T(Q)$.

Up to mass dimension two, only operators proportional to unity contribute to physical matrix elements of \ac{qcd}.
Explicitly, they read
\begin{equation}
  \calo^{(0)}_{1} \equiv \calo^{(0)} = \mathbb{1} ,\qquad \calo^{(2)}_{1} \equiv \calo^{(2)} = m_\bare^2\,\mathbb{1} ,
  \label{eq:dim02}
\end{equation}
where $m_\bare$ is the bare mass of the $\nh$ degenerate massive quarks.
Therefore, the Wilson coefficients
\begin{equation}
  \begin{split}
    C_1^{(0)}\equiv C^{(0)}\equiv C^{(0),\bare} , \qquad C_1^{(2)}\equiv C^{(2)}\equiv Z_m^2C^{(2),\bare}
  \end{split}
  \label{eq:dim2coefs}
\end{equation}
are \ac{uv}-finite, where $Z_m$ is the \msbar{} renormalization constant of the quark mass.
At mass dimension four we choose
\begin{equation}
  \calo^{(4)}_{1} \equiv \calo_{1} = \frac{1}{g_\bare^2}F^a_{\mu\nu} F^a_{\mu\nu} , \qquad
  \calo^{(4)}_{2} \equiv \calo_{2} = \sum_{f=1}^{\nf}\bar\psi_f\overleftrightarrow{\slashed{D}}^\mathrm{F} \psi_f , \qquad
  \calo^{(4)}_{3}\equiv \calo_{3} = m_\bare^4\,\mathbb{1}
  \label{eq:hvp-calo}
\end{equation}
as basis of operators.
Higher dimensional operators are neglected in the following.

Matrix elements of the dimension-four operators are divergent in general.
However, by defining renormalized operators $\calo_{n}^\ren$ as linear combinations among them, physical matrix elements as well as the Wilson coefficients become finite, i.e.
\begin{equation}
  \begin{split}
    \calo^\ren_{n} = \sum_k Z_{nk}\,\calo_{k} , \qquad  C_n = \sum_{m}C^\bare_m (Z^{-1})_{mn} ,
  \end{split}
  \label{eq:hvp-calob}
\end{equation}
where $C^\bare_n\equiv C_n^{(4),\bare}$, cf.\ \cref{eq:ccope,eq:hvp-calo}.
Since the operators of \cref{eq:hvp-calo} are part of the \ac{qcd} Lagrangian, the renormalization matrix $Z$ can be expressed in terms of the anomalous dimensions of \ac{qcd}~\cite{Spiridonov:1984br,Spiridonov:1988md}.

To derive the flowed \ac{ope}, we introduce the flowed operators as
\begin{equation}
  \begin{gathered}
    \tcalo_{1}(t,x) = \frac{Z_\mathrm{s}}{g_\bare^2}\,G^a_{\mu\nu}(t,x)G^a_{\mu\nu}(t,x) = \frac{4}{\hat{\mu}^{2\epsilon}g^2} E(t,x) ,\\
    \tcalo_{2}(t,x) = \mathring{Z}_\chi \sum_{f=1}^{\nf}\bar\chi_f(t,x) \overleftrightarrow{\slashed{\mathcal{D}}}^\mathrm{F} (t,x)\chi_f(t,x) = \mathring{R}(t,x) , \qquad
    \tcalo_{3}(t,x) = m^4\,\mathbb{1}  ,
  \end{gathered}
  \label{eq:flopo}
\end{equation}
where $E(t,x)$ and $\mathring{R}(t,x)$ are the composite operators already introduced in \cref{sec:VEVs}.
The small-flow-time expansion in \cref{eq:small-flow-time-expansion} allows us to relate the regular \ac{qcd} operators and coefficients with their flowed counterparts through
\begin{equation}
  \tcalo_{n}(t) = \zeta^{(0)}_{n}(t)\mathbb{1} + \zeta^{(2)}_{n}(t)\,m^2\mathbb{1} + \sum_{k}\zeta_{nk}(t)\calo^\ren_{k} + O(t) ,
  \label{eq:smalltexp}
\end{equation}
where $\zeta^{(2)}_{n}(t)$ and $\zeta_{nk}(t)$ are the renormalized, finite mixing coefficients.
Inverting \cref{eq:smalltexp} yields
\begin{equation}
    \calo^\ren_{n} = \sum_k \zeta_{nk}^{-1}(t)\,\bcalo_{k}(t) + O(t) , \qquad
    \bcalo_{n}(t) \equiv \tcalo_{n}(t) - \zeta^{(0)}_{n}(t)\mathbb{1} - \zeta^{(2)}_{n}(t)m^2\mathbb{1} ,
  \label{eq:opoinv}
\end{equation}
which leads to the flowed \ac{ope} for the current correlator:
\begin{equation}
  T(Q)\stackrel{Q^2\to\infty}{\sim} \tilde{C}^{(0)}(Q^2,t) + \tilde{C}^{(2)}(Q^2,t)m^2 + \sum_n\tilde{C}_{n}(Q^2,t)\langle\tcalo_{n}(t)\rangle + O(t) .
  \label{eq:flowedope}
\end{equation}
The flowed Wilson coefficients are related to the regular Wilson coefficients through
\begin{equation}
    \tilde C_{n}(Q^2,t) = \sum_k C_{k}(Q^2)\zeta_{kn}^{-1}(t) , \qquad
    \tilde C^{(0,2)}(Q^2,t) = C^{(0,2)}(Q^2) - \sum_{n}\tilde C_{n}(Q^2,t)\,\zeta^{(0,2)}_n(t) .
  \label{eq:ctilde}
\end{equation}
The regular Wilson coefficients $C^{(0)}$ and $C^{(2)}$ are given by the first two terms in $\nicefrac{m^2}{Q^2}$ of the large-$Q^2$ expansion of the \acp{vpf}.
Through the required order, they can be found in \citere{Chetyrkin:1997qi} for vector-, in \citere{Harlander:1997kw} for axial-vector-, and in \citere{Harlander:1997xa} for scalar- and pseudo-scalar currents, for example.
The dimension-four coefficients can be found in \citeres{Chetyrkin:1985kn,Harlander:diss}.\footnote{Since the latter reference is only available in German, they have also been included in \citere{Harlander:2020duo}.}

In \citere{Harlander:2020duo} we determined the mixing matrix $\zeta$ through \ac{nnlo} with the help of the method of projectors~\cite{Gorishnii:1983su,Gorishnii:1986gn}.
Most of the matrix elements can already be extracted from the calculation of the \acp{vev} in \cref{sec:VEVs} (or rather \citere{Artz:2019bpr}) as well as from the calculation of the \ac{emt} in \citere{Harlander:2018zpi}.
The remaining elements correspond to higher-order corrections in the bare mass to the \acp{vev}.
By combining $\zeta$ with the known results for the regular Wilson coefficients, one can determine the flowed coefficients of \cref{eq:ctilde} to the same order.
Together with an evaluation of the flowed operator matrix elements on the lattice, the \acp{vpf} can be extracted and used in the determination of various physical quantities.

\section{Conclusion}
In this contribution we discussed the perturbative gradient flow and stressed its powerful renormalization properties.
Then, we outlined the calculation of some \acp{vev} of gauge-invariant operators at finite flow time which enable the construction of a gradient flow coupling.
Afterwards, we discussed the flowed \ac{ope} which can be used to replace regular operators by their better behaved flowed counterparts and applied it to \acp{vpf} which might lead to new results for quantities like anomalous magnetic moments from lattice simulations.

\section*{Acknowledgements}
We thank Johannes Artz, Robert V.~Harlander, Tobias Neumann, and Mario Prausa for their collaboration on the projects presented in this contribution.
Furthermore, we thank Robert V.~Harlander and Tobias Neumann for comments on the manuscript.

\paragraph{Funding information}
The research projects presented in this contribution were supported by the \textit{Deutsche Forschungsgemeinschaft} (DFG, German Research Foundation) through grant \href{http://gepris.dfg.de/gepris/projekt/386986591?language=en}{386986591}.
Some of the calculations in \cref{sec:VEVs} were performed with computing resources granted by RWTH Aachen University under project ``rwth$0244$''.

\begin{appendix}
\end{appendix}

\bibliography{bib}

\nolinenumbers

\end{document}